\documentclass[review,number]{elsarticle}

\usepackage[utf8]{inputenc}
\usepackage[T1]{fontenc}
\usepackage{amsmath,amssymb}
\usepackage{graphicx}
\usepackage{listings}
\usepackage{xcolor}
\usepackage{booktabs}
\usepackage{hyperref}

\journal{Astronomy and Computing}

\begin{document}

\begin{frontmatter}

\title{A Unified HI Rotation Curve Corpus for Computational Astrophysics:\\438 Galaxies from SPARC, THINGS, LITTLE THINGS, and WALLABY DR2}

\author[eps]{David C. Flynn\corref{cor1}}
\ead{davidflynn@eps-research.com}
\cortext[cor1]{Corresponding author. ORCID: 0000-0002-2768-6650}
\address[eps]{EPS Research, Laurel MD, USA}

\begin{abstract}
We present a unified corpus of 8,963 spatially resolved HI rotation curve measurements across 423 galaxies (438 total catalog entries including 15 metadata-only THINGS galaxies), drawn from four major surveys: SPARC (175), THINGS (34), LITTLE THINGS (26), and WALLABY DR2 (203). The corpus is distributed as a single structured JSON file with nested per-ring kinematic data, survey metadata, column definitions, and data-quality annotations, accompanied by a 438-row flat CSV for catalog-level filtering. All radii are in kiloparsecs, all velocities in km/s. Kinematic parameters have been verified against scanned primary tables. A two-tier quality system distinguishes hand-curated rotation curves with per-point uncertainties (Tier~1) from automated pipeline products (Tier~2). The corpus was designed for both traditional numerical analysis and Large Language Model retrieval-augmented generation (RAG) pipelines. Three worked examples demonstrate single-galaxy rotation curve plotting, multi-component baryonic analysis, and corpus-level parameter-space exploration, each requiring fewer than 15 lines of Python. The corpus is publicly available at Zenodo (DOI: 10.5281/zenodo.19563417) under CC BY 4.0.
\end{abstract}


\end{frontmatter}


\section{Introduction}

Galaxy rotation curves remain the primary observational evidence for the mass discrepancy problem in disk galaxies. The gap between observed circular velocities and those predicted from visible baryonic matter has motivated dark matter halo models~\cite{navarro1996}, modified gravity theories~\cite{milgrom1983,mcgaugh2004}, and empirical correction frameworks~\cite{flynn2025}. All of these approaches require access to high-quality, spatially resolved rotation curve data with consistent units and metadata.

Four major HI surveys currently provide the bulk of published rotation curves for nearby galaxies. SPARC (Spitzer Photometry and Accurate Rotation Curves~\cite{lelli2016}) offers 175 galaxies with full baryonic decomposition at 3.6~$\mu$m. THINGS (The HI Nearby Galaxy Survey~\cite{walter2008,deblok2008}) provides high-resolution VLA tilted-ring fits for 19 galaxies. LITTLE THINGS~\cite{oh2015} contributes 26 dwarf irregular galaxies observed at comparable VLA resolution. WALLABY DR2 (Widefield ASKAP L-band Legacy All-sky Blind Survey~\cite{westmeier2022,deg2022,murugeshan2024}) adds 203 galaxies from the ASKAP automated pipeline, extending coverage to previously uncharacterized systems.

Despite the scientific importance of these datasets, no unified, machine-readable corpus currently exists that combines all four surveys in a single schema. While the individual survey products remain publicly available, they are distributed across heterogeneous platforms with incompatible formats, differing column conventions (arcseconds vs.\ kiloparsecs, varying velocity definitions), and no common schema---conditions that impose a significant barrier to reproducible computational analysis and effectively render the combined dataset inaccessible to automated pipelines. Researchers wishing to compare rotation curves across surveys must individually download these disparate source files, reconcile unit and naming conventions, and manually verify kinematic parameters against scattered primary publications. This data-fragmentation problem is particularly acute for computational workflows---including automated fitting pipelines, statistical meta-analyses, and emerging LLM-based retrieval-augmented generation (RAG) architectures---that require structured, self-describing input data. The problem is likely to intensify as earlier-generation HI surveys hosted on aging or deprecated infrastructure become candidates for similar standardization efforts.

This paper presents the \emph{Unified Galaxy HI Rotation Curve Corpus} (v7.0), a structured JSON dataset containing 8,963 individually resolved rotation curve measurements across 423 galaxies from all four surveys, with an additional 15 THINGS galaxies contributing verified kinematic metadata. We describe the corpus architecture (Section~3), present the catalog-level CSV schema (Section~4), document the ingestion and verification procedures (Section~5), demonstrate three concrete usage examples with embedded figures (Section~6), discuss the LLM/RAG application (Section~7), and enumerate known limitations (Section~8). The corpus is publicly available at Zenodo (DOI: 10.5281/zenodo.19563417) under CC BY 4.0.


\section{Survey Coverage and Source Data}

Table~\ref{tab:coverage} summarizes the survey coverage. The corpus contains rotation curve data for 423 galaxies and kinematic metadata for a further 15, totaling 438 catalog entries and 8,963 radial measurement points.

\begin{table}[htbp]
\caption{Survey coverage summary.}
\label{tab:coverage}
\centering
\begin{tabular}{lrrll}
\toprule
Survey & Galaxies & Data Points & Tier & Primary Reference \\
\midrule
SPARC         & 175           & 3,391 & 1 & Lelli et al.\ (2016)~\cite{lelli2016} \\
THINGS        & 34 (19 w/data)& 2,110 & 1 & de Blok et al.\ (2008)~\cite{deblok2008} \\
LITTLE THINGS & 26            & 1,716 & 1 & Oh et al.\ (2015)~\cite{oh2015} \\
WALLABY DR2   & 203           & 1,746 & 2 & Deg et al.\ (2022)~\cite{deg2022}; \\
              &               &       &   & Murugeshan et al.\ (2024)~\cite{murugeshan2024} \\
\midrule
\textbf{Total} & \textbf{438} & \textbf{8,963} & & \\
\bottomrule
\end{tabular}
\end{table}

\subsection{SPARC}

The SPARC database~\cite{lelli2016} provides 175 galaxies with HI/H$\alpha$ rotation curves and full baryonic decomposition at Spitzer 3.6~$\mu$m. Each galaxy carries observed velocity ($V_\mathrm{obs}$), gas velocity ($V_\mathrm{gas}$), disk velocity ($V_\mathrm{disk}$), bulge velocity ($V_\mathrm{bul}$), and surface brightness profiles at each radial point, enabling mass modeling without external photometry. Kinematic parameters (inclination, distance, and their uncertainties) were verified against Lelli et al.~\cite{lelli2016} Table~1 (Lelli2016c.mrt; scanned). SPARC coordinates (RA, Dec), systemic velocities, and position angles are absent from the corpus because they were not published in a unified SPARC table and remain distributed across approximately 50 individual source papers.

\subsection{THINGS}

The THINGS survey~\cite{walter2008} observed 34 nearby galaxies at high spectral and spatial resolution with the VLA. De Blok et al.~\cite{deblok2008} published tilted-ring rotation curves for 19 of these galaxies. The remaining 15 were excluded from the kinematic analysis due to morphological disturbance, low inclination, strong non-circular motions, or other factors. These 15 galaxies are included in the corpus with verified metadata (distance, inclination, position angle, coordinates) but without per-point rotation curve data. Fourteen also appear in SPARC and carry full baryonic decomposition under their SPARC entries. Kinematic parameters were verified against de Blok et al.~\cite{deblok2008} Tables~1 and~2 (scanned).

\subsection{LITTLE THINGS}

LITTLE THINGS (Local Irregulars That Trace Luminosity Extremes, The HI Nearby Galaxy Survey~\cite{oh2015}) contributes 26 dwarf irregular galaxies observed with the VLA, providing rotation velocities ($V_\mathrm{rot}$) with per-point uncertainties. All kinematic parameters were verified against Oh et al.~\cite{oh2015} Table~1 (scanned). The characteristic radius $r_{0.3}$ and corresponding velocity $v_{0.3}$ (the radius at which the logarithmic slope of the rotation curve equals 0.3) are included in the catalog CSV.

\subsection{WALLABY DR2}

The WALLABY DR2 kinematic catalogue~\cite{deg2022,murugeshan2024} provides spatially resolved rotation curves for 203 galaxies processed through the WALLABY Kinematic Analysis Pipeline Products (WKAPP) automated pipeline: the 3DBarolo tilted-ring fitter~\cite{diteodoro2015} combined with the Fully Automated TiRiFiC (FAT) pipeline~\cite{kamphuis2015}. The ASKAP synthesised beam is 30 arcseconds (FWHM), setting a practical spatial resolution floor. $V_\mathrm{rot}$ values below 50~km/s are subject to beam-smearing and should be treated with caution. Distances are derived from Hubble flow at $H_0 = 75~\mathrm{km\,s^{-1}\,Mpc^{-1}}$ unless Cosmicflows-4 distances are available. Per-ring rotation velocity uncertainties are not published in the DR2 catalogue; WALLABY entries therefore carry \texttt{quality\_tier}~=~2. Of 203 WALLABY galaxies, 202 carry quality flag~0 (reliable model fit) and 1 carries flag~1 (marginal). Ring counts range from 3 to 47 (median~7). Per-ring data include radius (kpc), $V_\mathrm{rot}$ (km/s), velocity dispersion (km/s), inclination (degrees), and position angle (degrees).


\section{Corpus Architecture and Schema}

\subsection{File formats}
\label{sec:formats}

The corpus is distributed in three complementary formats. The master file \emph{rotation\_curve\_corpus\_v7.json} is a single JSON document ($\sim$2.0~MB) containing all 438 galaxy entries within a unified schema, together with a top-level metadata block encoding version, survey counts, quality tier definitions, and citation. The flat table \emph{rotation\_curve\_corpus\_v7\_flat.csv} provides one row per galaxy (438 rows, 29 columns) with summary statistics for sample selection and filtering. The per-galaxy archive \emph{rotation\_curve\_corpus\_v7\_by\_galaxy.zip} contains 438 individual JSON files organised into subdirectories by survey (SPARC/, THINGS/, LITTLE\_THINGS/, WALLABY/), each self-contained with full corpus metadata and the complete rotation curve array, optimised for LLM/RAG ingestion where each galaxy constitutes a single retrieval document.

\subsection{Quality tier system}

A two-tier quality annotation is applied at the galaxy level. Tier~1 (SPARC, THINGS, LITTLE THINGS) denotes hand-curated rotation curves with per-point uncertainties, verified kinematic parameters, and---for SPARC---full baryonic decomposition. Tier~2 (WALLABY DR2) denotes automated pipeline products from the WKAPP system, peer-reviewed but without per-ring uncertainties or baryonic components. The tier system enables downstream analyses to filter by data provenance without inspecting individual galaxies.

\subsection{Units and conventions}

All radii are stored in kiloparsecs. THINGS radii were converted from arcseconds using $R[\mathrm{kpc}] = R[\mathrm{arcsec}] \times D[\mathrm{Mpc}] \times 1000 \times \pi / 648000$. All velocities are in km/s. SPARC baryonic velocity components ($V_\mathrm{gas}$, $V_\mathrm{disk}$, $V_\mathrm{bul}$) are stored at mass-to-light ratio $\Upsilon = 1$. The gas velocity $V_\mathrm{gas}$ may be negative at inner radii due to the sign-preserving quadrature convention used throughout SPARC, in which thermal pressure exceeds rotational support. The total baryonic velocity is computed via:
\begin{equation}
V_\mathrm{bar} = \sqrt{\Upsilon_\star V_\mathrm{disk}^2 + \Upsilon_b V_\mathrm{bul}^2 + \mathrm{sign}(V_\mathrm{gas}) \, V_\mathrm{gas}^2}
\end{equation}
The \texttt{vgas\_negative\_rows} column in the CSV records the count of inner-disk radii where $V_\mathrm{gas} < 0$ for each SPARC galaxy. All tables presented in this manuscript are rounded to instrument-appropriate precision; the machine-readable JSON and CSV files retain full floating-point values to preserve round-trip fidelity for programmatic use (see Section~8, Limitation~5).

\subsection{JSON schema by survey}

Because the four surveys provide fundamentally different observables, the per-galaxy JSON schema varies by survey. Table~\ref{tab:schema} summarizes the per-ring columns available for each.

\begin{table}[htbp]
\caption{Per-ring data columns by survey. SPARC/THINGS/LITTLE THINGS use the \texttt{data} key; WALLABY uses \texttt{rotation\_curve}. Note that within the \texttt{data} key, SPARC and LITTLE THINGS use \texttt{Vobs}/\texttt{errV} while THINGS uses \texttt{Vrot}/\texttt{e\_Vrot}.}
\label{tab:schema}
\centering
\begin{tabular}{lll}
\toprule
Survey & Data Key & Per-Ring Columns \\
\midrule
SPARC (175) & \texttt{data} & Rad, Vobs, errV, Vgas, Vdisk, Vbul, SBdisk, SBbul \\
THINGS (19) & \texttt{data} & Rad, Vrot, e\_Vrot \\
LITTLE THINGS (26) & \texttt{data} & Rad, Vobs, errV \\
WALLABY (203) & \texttt{rotation\_curve} & rad\_kpc, vrot\_kms, vdisp\_kms, inc\_deg, pa\_deg \\
\bottomrule
\end{tabular}
\end{table}

Code accessing per-point data across surveys should check for both keys:
\begin{lstlisting}
points = galaxy.get('data') or galaxy.get('rotation_curve', [])
\end{lstlisting}


\section{Catalog-Level CSV Schema}

The flat CSV contains 29 columns per galaxy. Table~\ref{tab:csv} lists each field with its survey coverage. The CSV is derived from the JSON and is designed for rapid sample selection; per-point data require loading the JSON.

\begin{table}[htbp]
\caption{Flat CSV schema (29 columns). LT = LITTLE THINGS; W = WALLABY.}
\label{tab:csv}
\centering
\small
\begin{tabular}{llll}
\toprule
Field & Surveys & Coverage & Description \\
\midrule
\texttt{galaxy}            & All         & 438/438 & Galaxy identifier \\
\texttt{survey}            & All         & 438/438 & SPARC / THINGS / LITTLE\_THINGS / WALLABY \\
\texttt{quality\_tier}     & All         & 438/438 & 1 = hand-curated, 2 = automated \\
\texttt{telescope}         & THINGS,LT,W & 263/438 & Instrument identifier \\
\texttt{ra\_deg}           & THINGS,LT,W & 263/438 & Right ascension (J2000, deg) \\
\texttt{dec\_deg}          & THINGS,LT,W & 263/438 & Declination (J2000, deg) \\
\texttt{distance\_mpc}     & All         & 438/438 & Distance (Mpc) \\
\texttt{e\_distance\_mpc}  & SPARC       & 175/438 & Distance uncertainty (Mpc) \\
\texttt{vsys\_kms}         & THINGS,LT,W & 263/438 & Systemic velocity (km/s) \\
\texttt{inc\_deg}          & All         & 438/438 & Inclination (deg); verified \\
\texttt{e\_inc\_deg}       & SPARC       & 175/438 & Inclination uncertainty (deg) \\
\texttt{pa\_deg}           & THINGS,LT,W & 261/438 & Position angle (deg) \\
\texttt{hubble\_type}      & SPARC       & 175/438 & Hubble type integer (0--11) \\
\texttt{m2l\_disk}         & SPARC       & 175/438 & Mass-to-light ratio at [3.6] \\
\texttt{n\_points}         & All         & 423/438 & Rotation curve point count \\
\texttt{r\_min\_kpc}       & All         & 423/438 & Innermost radius (kpc) \\
\texttt{r\_max\_kpc}       & All         & 423/438 & Outermost radius (kpc) \\
\texttt{vrot\_min\_kms}    & All         & 423/438 & Min rotation velocity (km/s) \\
\texttt{vrot\_mean\_kms}   & All         & 423/438 & Mean rotation velocity (km/s) \\
\texttt{vrot\_max\_kms}    & All         & 423/438 & Peak rotation velocity (km/s) \\
\texttt{vdisp\_mean\_kms}  & WALLABY     & 203/438 & Mean velocity dispersion (km/s) \\
\texttt{has\_bulge}        & SPARC       & 201/438 & Boolean bulge flag \\
\texttt{vgas\_negative\_rows} & SPARC    & 175/438 & Inner radii with $V_\mathrm{gas} < 0$ \\
\texttt{r0p3\_kpc}         & LT          & 26/438  & Radius at log slope = 0.3 (kpc) \\
\texttt{v0p3\_kms}         & LT          & 26/438  & Velocity at $r_{0.3}$ (km/s) \\
\texttt{beam\_arcsec}      & WALLABY     & 203/438 & Beam FWHM (arcsec) \\
\texttt{qflag\_model}      & WALLABY     & 203/438 & 3DBarolo quality flag \\
\texttt{reference}         & All         & 438/438 & Primary citation \\
\texttt{notes}             & All         & 41/438  & Data quality notes \\
\bottomrule
\end{tabular}
\end{table}


\section{Data Ingestion and Verification}

\subsection{SPARC, THINGS, and LITTLE THINGS}

SPARC rotation curves were ingested from the publicly available flat files at astroweb.cwru.edu/SPARC. THINGS rotation curves were digitised from de Blok et al.~\cite{deblok2008} Table~2; radii were converted from arcseconds to kiloparsecs using the published distances. LITTLE THINGS data were obtained from VizieR (J/AJ/149/180). All kinematic parameters were cross-verified against scanned primary tables: Lelli et al.~\cite{lelli2016} Table~1 for SPARC, de Blok et al.~\cite{deblok2008} Tables~1 and~2 for THINGS, and Oh et al.~\cite{oh2015} Table~1 for LITTLE THINGS. This verification step caught an arcsec-to-kpc conversion bug (a missing $\times 1000$ factor) in an earlier ingestion iteration, underscoring the value of systematic primary-source checking.

\subsection{WALLABY DR2}

WALLABY rotation curves were ingested from the published 3DBarolo model files available at the Canadian Astronomy Data Centre (CADC). A Python script (\emph{wallaby\_ingest.py}, included as supplementary material) reads the CADC per-galaxy model output, extracts per-ring kinematic parameters, and cross-matches against the WALLABY DR2 kinematic catalogue (303 entries) to attach galaxy-level metadata (coordinates, distance, systemic velocity, inclination, position angle). Of 303 catalogue entries, 204 had Barolo model files at CADC; after quality screening, 203 were ingested. Ring counts range from 3 to 47 (median~7), with a total of 1,746 radial measurement points.

\subsection{Cross-matching}

Fourteen galaxies appear in two surveys (primarily SPARC and THINGS). The crossmatch index is stored in the JSON metadata block and enables downstream analyses to identify duplicate entries, apply survey-specific treatments, or merge complementary data (e.g., SPARC baryonic decomposition with THINGS tilted-ring kinematics for the same galaxy).


\section{Usage Examples}

The following three examples demonstrate the corpus's utility for common rotation curve analyses. Each example loads data directly from the JSON with no external preprocessing. All code is Python~3 using only the standard library, numpy, and matplotlib. The figure-generation script (\emph{make\_figures\_v7.py}) is included as supplementary material.

\subsection{Example 1: Multi-component baryonic analysis (SPARC)}

\begin{figure}[htbp]
\centering
\includegraphics[width=\textwidth]{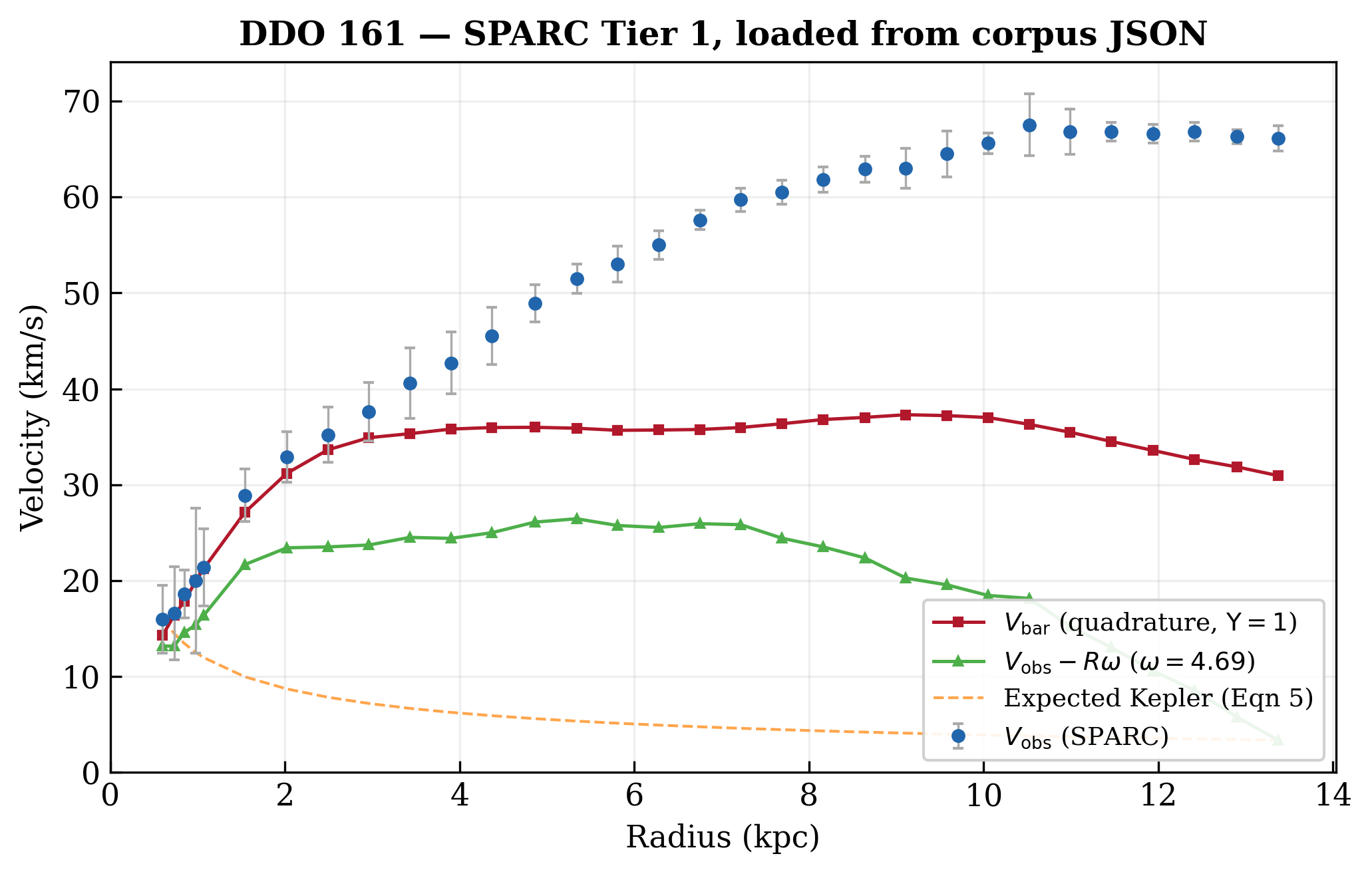}
\caption{DDO~161 (SPARC Tier~1) loaded from corpus JSON. Blue circles: $V_\mathrm{obs}$ with SPARC error bars. Red squares: $V_\mathrm{bar}$ from sign-preserving quadrature at $\Upsilon = 1$. Green triangles: $V_\mathrm{obs} - R\omega$ ($\omega = 4.69$~rad/Gyr~\cite{flynn2025}). Orange dashed: expected Keplerian baseline.}
\label{fig:ddo161}
\end{figure}

Figure~\ref{fig:ddo161} shows DDO~161 (a SPARC Tier~1 dwarf irregular galaxy at 7.4~Mpc) with four curves extracted directly from the corpus JSON: $V_\mathrm{obs}$ with SPARC error bars (blue circles), $V_\mathrm{bar}$ from sign-preserving quadrature at $\Upsilon = 1$ (red squares), the omega-corrected velocity $V_\mathrm{obs} - R\omega$ (green triangles, applying the empirical correction $V_\mathrm{obs} = V_\mathrm{Kepler} + R\omega$, where $\omega$ is a per-galaxy angular velocity offset~\cite{flynn2025}, here $\omega = 4.69$~rad/Gyr; note that 1~rad/Gyr $\approx$ 1.022~km/s/kpc, so $\omega$ in these units yields $V$ in km/s when multiplied by $R$ in kpc), and the expected Keplerian baseline (orange dashed). The gap between blue and red is the mass discrepancy that dark matter, MOND, or the omega correction each attempt to explain. The point for this data descriptor is that all four curves come from a single JSON load with no preprocessing.

\begin{lstlisting}
import json, numpy as np, matplotlib.pyplot as plt

with open('rotation_curve_corpus_v7.json') as f:
    corpus = json.load(f)
g = next(g for g in corpus['galaxies'] if g['galaxy']=='DDO161')
d = g['data']
R    = np.array([p['Rad']  for p in d])
Vobs = np.array([p['Vobs'] for p in d])
errV = np.array([p['errV'] for p in d])
Vgas = np.array([p['Vgas'] for p in d])
Vdisk= np.array([p['Vdisk'] for p in d])
Vbul = np.array([p['Vbul'] for p in d])

# Baryonic velocity: sign-preserving quadrature
Vbar = np.where(Vgas<0, -np.sqrt(Vgas**2+Vdisk**2+Vbul**2),
                         np.sqrt(Vgas**2+Vdisk**2+Vbul**2))
omega = 4.69  # rad/Gyr, [4] Table 2
V_omega = Vobs - R * omega
\end{lstlisting}

\subsection{Example 2: WALLABY rotation curve with caution zone (Tier 2)}

\begin{figure}[htbp]
\centering
\includegraphics[width=0.75\textwidth]{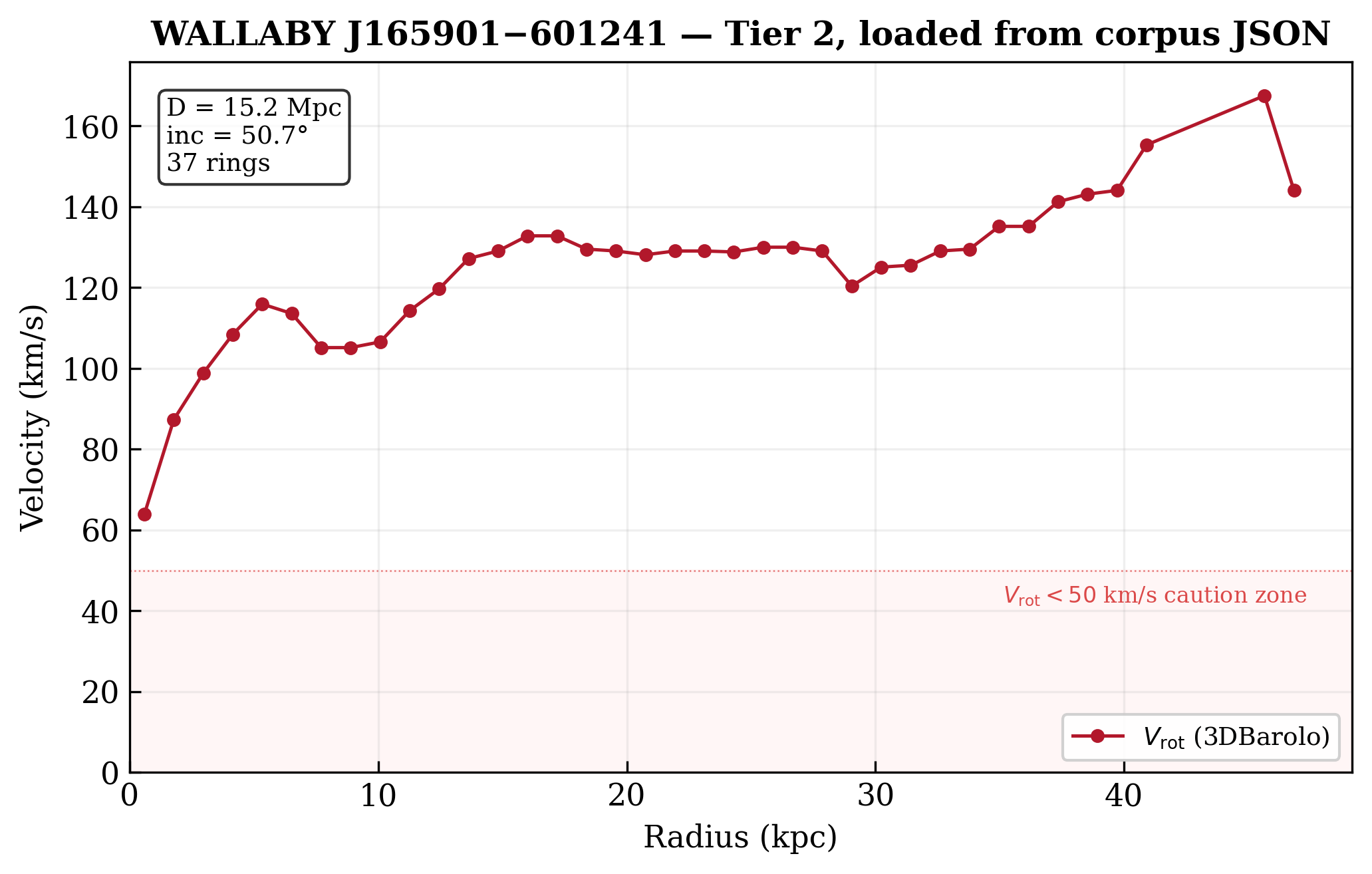}
\caption{WALLABY J165901$-$601241 (Tier~2) loaded from corpus JSON. The 50~km/s beam-smearing caution zone is shaded. Metadata ($D = 15.2$~Mpc, inc $= 50.7^\circ$, 37 rings) is extracted directly from the JSON.}
\label{fig:wallaby}
\end{figure}

Figure~\ref{fig:wallaby} demonstrates the Tier~2 pipeline output for WALLABY J165901$-$601241, a galaxy with 37 rings and a clean rising curve to $\sim$167~km/s. All points lie well above the 50~km/s beam-smearing caution zone (shaded). The metadata annotation (distance, inclination, ring count) is extracted directly from the JSON, showing the corpus carries everything needed for quality assessment.

\begin{lstlisting}
wg = next(g for g in corpus['galaxies']
          if g.get('galaxy')=='WALLABY_J165901-601241')
rc = wg['rotation_curve']
R_w = [p['rad_kpc']  for p in rc]
V_w = [p['vrot_kms'] for p in rc]

plt.plot(R_w, V_w, 'o-', color='#B2182B')
plt.axhspan(0, 50, alpha=0.15, color='red',
            label='Vrot < 50 km/s (caution)')
plt.text(0.03, 0.95, f"D={wg['distance_mpc']:.1f} Mpc\n"
         f"inc={wg['inc_deg']:.1f} deg\n{len(rc)} rings",
         transform=plt.gca().transAxes, va='top')
\end{lstlisting}

\subsection{Example 3: Corpus-level parameter-space exploration}

\begin{figure}[htbp]
\centering
\includegraphics[width=\textwidth]{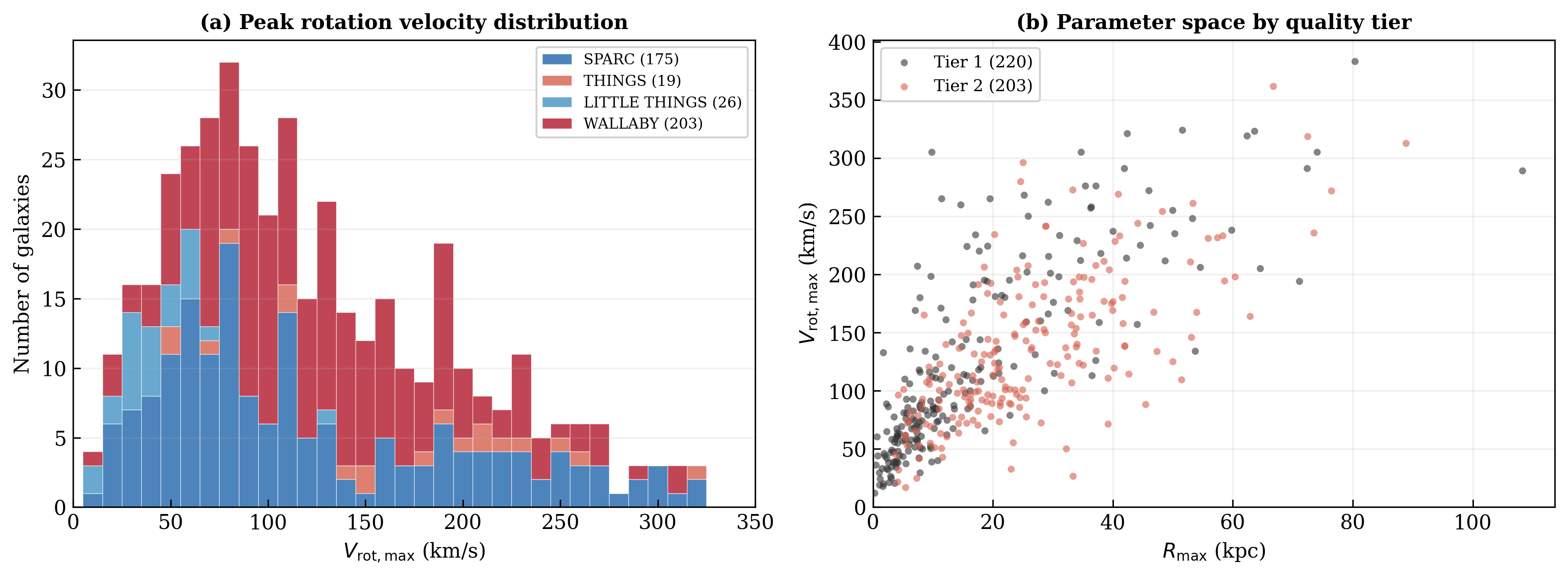}
\caption{Corpus population overview. (a)~Peak rotation velocity distribution across all four surveys (stacked histogram, $N = 423$ galaxies with rotation curve data). (b)~$R_\mathrm{max}$ vs.\ $V_\mathrm{rot,max}$ parameter space colored by quality tier (Tier~1: 220 galaxies, black; Tier~2: 203 galaxies, red). The corpus spans from $\sim$3~kpc dwarf irregulars to $\sim$100~kpc massive spirals.}
\label{fig:population}
\end{figure}

Figure~\ref{fig:population} provides the corpus-level overview. Panel~(a) shows the peak rotation velocity distribution across all four surveys as a stacked histogram: SPARC and WALLABY dominate different velocity ranges, while THINGS and LITTLE THINGS fill intermediate coverage. Panel~(b) shows the $R_\mathrm{max}$ vs.\ $V_\mathrm{rot,max}$ parameter space colored by quality tier, demonstrating that the corpus spans from $\sim$3~kpc dwarf irregulars to $\sim$100~kpc massive spirals.

\begin{lstlisting}
# Panel (a): stacked histogram by survey
for survey in ['SPARC','THINGS','LITTLE_THINGS','WALLABY']:
    vals = [max(p.get('Vobs',p.get('vrot_kms',0))
            for p in g.get('data') or g.get('rotation_curve',[]))
            for g in corpus['galaxies'] if g['survey']==survey
            and (g.get('data') or g.get('rotation_curve'))]
    ax[0].hist(vals, bins=36, range=(0,350),
              label=f'{survey} ({len(vals)})', alpha=0.8)

# Panel (b): scatter by tier
ax[1].scatter(r_max[tier==1], v_max[tier==1], label='Tier 1')
ax[1].scatter(r_max[tier==2], v_max[tier==2], label='Tier 2')
\end{lstlisting}

\section{Application to LLM-Based Inference}

A secondary design goal of the corpus is to serve as a retrieval corpus for LLM-based RAG pipelines in astrophysical research. In a RAG architecture, a user query (e.g., ``plot the rotation curve of DDO~161 and compute its baryonic mass model'') triggers a retrieval step that locates the relevant galaxy document, which is then injected into the LLM's context window alongside the query. The per-galaxy ZIP archive described in Section~\ref{sec:formats} is optimised for this use case: each file is a self-contained JSON document of $\sim$1--5~KB, well within typical context limits, containing all metadata and per-ring data needed to answer the query without external lookups.

To assess whether the JSON schema is sufficiently self-describing for automated consumption, we conducted a structured usability evaluation with four LLMs: Google Gemini Pro, Anthropic Claude, AstroSage-Llama-3.1-70B (a domain-specific astronomy foundation model), and Microsoft Copilot Pro. Each model was presented with a single per-galaxy JSON document and asked to perform three benchmark tasks without additional documentation: (1) plot the rotation curve with error bars, (2) compute the baryonic velocity via sign-preserving quadrature, and (3) apply the omega correction from~\cite{flynn2025}. All four models successfully generated syntactically correct Python scripts for all three tasks on first attempt, requiring no additional prompting beyond a natural-language research question. Gemini Pro and Claude produced the most fluent end-to-end workflows, generating matplotlib figures directly within their native interfaces. Copilot Pro generated correct Python but required script execution outside the chat environment to render figures. AstroSage-70B, tested via LM Studio, proved effective at identifying corpus-level defects and schema inconsistencies but was more sensitive to the host platform's capabilities; RAG-optimised deployment environments may yield stronger results. In all cases, the most reliable workflow was to request Python code generation and execute the resulting scripts in a Jupyter notebook, which enabled the full range of rotation curve analysis, baryonic decomposition, and omega correction application. These results suggest that the corpus's explicit column definitions, unit annotations, and quality flags provide sufficient context for LLM-based code generation and analysis without external documentation, and that the per-galaxy JSON format integrates naturally into interactive computational environments.


\section{Known Limitations}

Five limitations should be noted by users of this corpus.

\textbf{(1) Fifteen THINGS galaxies lack per-point rotation curve data.} As noted in Section~2.2, these galaxies were excluded from the de Blok et al.~\cite{deblok2008} tilted-ring analysis for legitimate astrophysical reasons and are marked with \texttt{n\_points = null} in the CSV.

\textbf{(2) SPARC entries lack coordinates and systemic velocities.} SPARC RA, Dec, $V_\mathrm{sys}$, and PA are absent because Lelli et al.~\cite{lelli2016} did not publish these in a unified table. These parameters are recoverable from NED or SIMBAD for individual galaxies; they were not included in the corpus to avoid introducing a secondary provenance layer beyond the original survey publications.

\textbf{(3) WALLABY rotation curves carry no per-ring uncertainties and no baryonic decomposition.} $V_\mathrm{rot}$ below 50~km/s is unreliable due to beam smearing at 30~arcsec ASKAP resolution.

\textbf{(4) JSON schema is not fully harmonised across surveys.} SPARC/THINGS/LITTLE THINGS use a \texttt{data} key; WALLABY uses \texttt{rotation\_curve} with different field names. This reflects genuine differences in the underlying observables.

\textbf{(5) Decimal precision may exceed measurement accuracy in some derived columns.} Floating-point artifacts from unit conversions (e.g., arcsec to kpc) are retained in the JSON to preserve round-trip fidelity. Radii are stored to three decimal places ($\sim$1~pc); velocities to two decimal places ($\sim$0.01~km/s). Users applying uncertainty-propagation analysis should round to instrumental precision before reporting results. Typical instrumental uncertainties are 1--5~km/s for rotation velocities and $\sim$0.1~kpc for radii at the distances of these surveys.


\section{Data Availability}

The corpus is publicly available at Zenodo under version-specific DOI: \textbf{10.5281/zenodo.19563417} (concept DOI for all versions: 10.5281/zenodo.19425427). The deposited files are named v7.0 to reflect internal development versioning; this Zenodo record represents the first peer-reviewed public release of the unified corpus. The deposit includes the master JSON, flat CSV, per-galaxy ZIP archive, corpus description sheet, and the WALLABY ingestion script. The corpus schema, normalisation, annotations, and unified structure are original work by D.C.\ Flynn / EPS Research and are released under CC BY 4.0. All underlying rotation curve data are drawn from published, publicly available sources; users should cite both this corpus and the relevant survey papers listed in the references.


\section*{Acknowledgements}

This work was conducted as independent research by EPS Research without external funding or institutional affiliation. The author thanks Jim Cannaliato for collaboration on the omega correction framework. The SPARC database is maintained by Federico Lelli and Stacy McGaugh at Case Western Reserve University. THINGS data products are based on observations with the Karl G.\ Jansky Very Large Array of the National Radio Astronomy Observatory. WALLABY data products are based on observations with the Australian Square Kilometre Array Pathfinder (ASKAP) and were accessed via the Canadian Astronomy Data Centre (CADC).


\section*{Declaration of Generative AI Use}

In accordance with Elsevier's policy on the use of generative AI and AI-assisted technologies, the author discloses the following. Four large language models were used during the creation and validation of this corpus and manuscript: Google Gemini Pro, Anthropic Claude (Opus and Sonnet), AstroSage-Llama-3.1-70B, and Microsoft Copilot Pro. Because the corpus is explicitly designed for LLM-based retrieval-augmented generation (RAG) pipelines, these models served as both development tools and validation instruments:

(1)~\emph{Corpus schema design and ingestion code.} LLMs assisted in drafting and debugging the Python ingestion scripts (including wallaby\_ingest.py), the JSON schema design, and the flat CSV generation code. All code was reviewed, tested, and validated by the author against primary source data.

(2)~\emph{Data verification and error detection.} LLMs were used to cross-check ingested values against scanned primary tables~\cite{lelli2016,deblok2008,oh2015}, which led to the detection and correction of an arcsec-to-kpc conversion bug in an earlier ingestion iteration.

(3)~\emph{RAG validation testing.} As described in Section~7, Gemini Pro, Claude, AstroSage-70B, and Copilot Pro were each tested as downstream consumers of the corpus JSON to verify that the schema is sufficiently self-describing for LLM-based scientific analysis---including rotation curve plotting, sign-preserving baryonic quadrature, and omega correction application---without additional prompting beyond natural-language research questions.

(4)~\emph{Manuscript preparation.} Claude (Anthropic) assisted in drafting, formatting, and assembling this manuscript, including figure generation code and document production. All scientific content, interpretations, data provenance decisions, and editorial judgments are the sole responsibility of the author.

No generative AI output was accepted without human review. The author takes full responsibility for the content of this publication.


\end{document}